\documentclass[aps,prb,twocolumn,groupedaddress]{revtex4}

\usepackage[dvips]{color}
\usepackage{graphicx,color}
\begin{document}

\title{Robustness of Majorana Fermion induced Fractional Josephson Effect}

\author{ K. T. Law, Patrick A. Lee}

\affiliation{Department of Physics, Massachusetts Institute of Technology, Cambridge, Massachusetts, 02139, USA}

\begin{abstract}
It is shown in previous works that the coupling between two Majorana end states in superconducting quantum wires leads to fractional Josephson effect. However, in realistic experimental conditions, multiple bands of the wires are occupied and the Majorana end states are accompanied by other fermionic end states. This raises the question concerning the robustness of fractional Josephson effect in these situations. In this work, we show that the absence of the avoided energy crossing which gives rise to the fractional Josephson effect is robust, even when the Majorana fermions are coupled with arbitrary strengths to other fermions. Moreover, we calculate the temperature dependence of the fractional Josephson current and show that it is suppressed by thermal excitations to the other fermion bound states.
\end{abstract}


\maketitle

\begin{figure}
\includegraphics[width=2.8in]{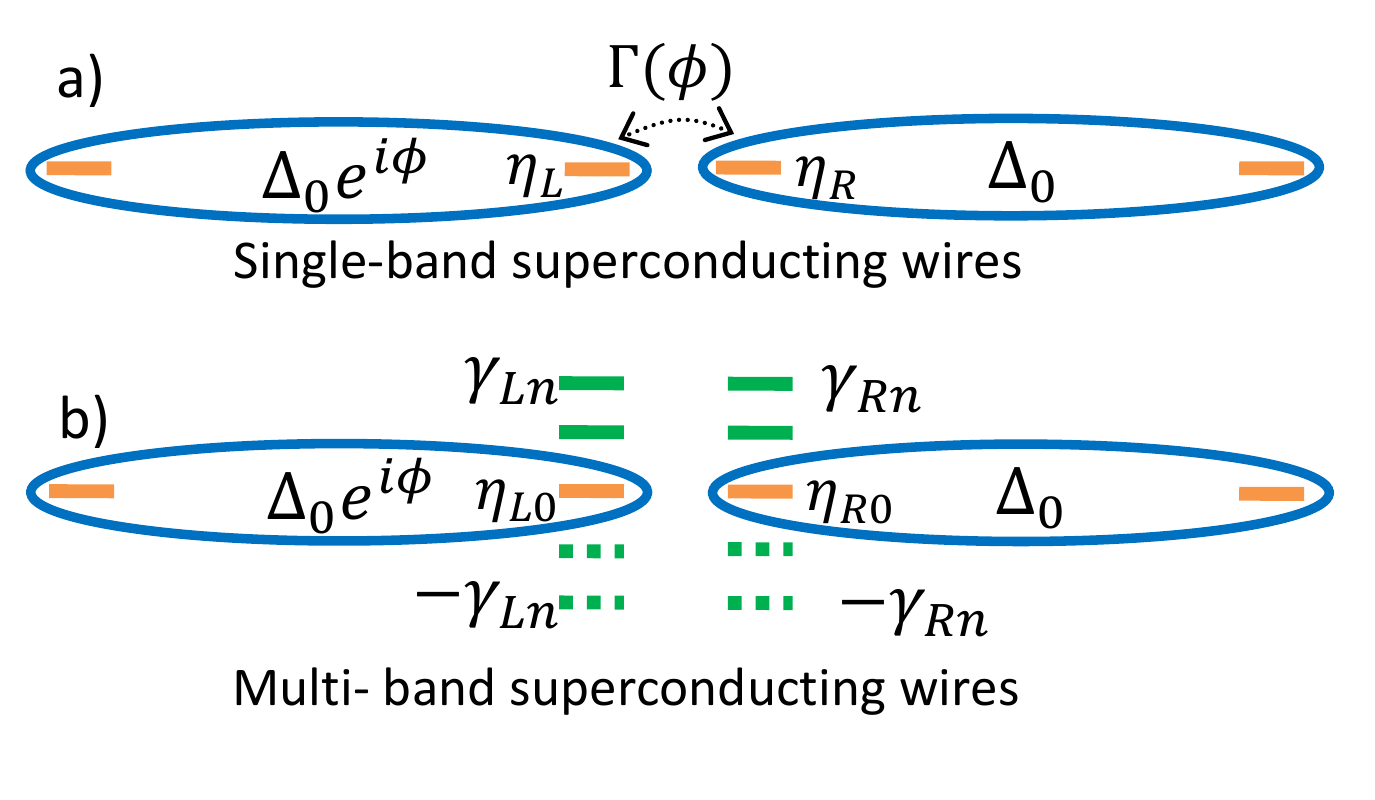}
\caption{\label{Fig1} a) Two 1D superconducting wires are brought together to form a Josephson junction. The Majorana end states at the junction are labeled by $\eta_{L}$ and $\eta_R$ respectively. These two Majorana end states at the junction couple to each other with amplitude $\Gamma(\phi)$ to form an Andreev bound state. The spectrum of the Andreev bound state is shown in Fig.\ref{Fig2}a. b) When multiple bands are occupied, the Majorana end states are accompanied by other fermionic states, denoted by $\gamma_{L/Rn}$ and $-\gamma_{L/Rn}$ (occupied and empty states respectively).}
\end{figure}

\emph{Introduction}--- Majorana fermions in condensed matter systems have drawn much attention recently because they obey non-Abelian statistics and have potential applications in quantum computation which are protected from local perturbations [\onlinecite{NSSFD, Ivanov}]. A number of systems are proposed to host Majorana fermions, including $p_{x}+ip_{y}$ superconductors [\onlinecite{RG}], topological insulator/superconductor heterostructures[\onlinecite{FKPRL08}], spin-orbit coupled semiconductor-superconductor heterostructure [\onlinecite{Fujimoto,SLTD,Alicea,Lee,PL11}] and metallic thin-films in proximity to superconductors[\onlinecite{PL}]. Majorana fermions in these systems manifest themselves through several interesting effects. For example, they induce cross-Andreev reflections[\onlinecite{NAB}], resonant Andreev reflection [\onlinecite{LLN}], electron teleportation [\onlinecite{Fu}] etc. Moreover, when two Majorana end states from two separate superconducting quantum wires are brought close to each other to form a Josephson junction, they induce fractional Josephson effect [\onlinecite{Kitaev00,KSY,FKPRB09,LSD,AOROF}]. The periodicity of the resulting Josephson current is $4\pi$ in terms of the phase difference between the two superconducting quantum wires, instead of $2\pi$ as in usual Josephson junctions. A schematic picture of the Josephson junction is shown in Fig.\ref{Fig1}a.

When two Majorana end states are brought close to each other to form the tunnel junction, they form a fermion state which may be occupied or empty. These two states form the q-bit in proposed quantum computation schemes. The fractional Josephson effect is important because its sign depends on the occupation of the fermion state and provides a direct way to read out the q-bit [\onlinecite{Kitaev00,KSY,FKPRB09,LSD,AOROF}].

However, in more realistic experimental conditions, multiple bands of the quantum wires are occupied and the Majorana end states are accompanied by other fermion end states. A schematic picture is shown in Fig.\ref{Fig1}b. These fermion end states may couple to the Majorana end states strongly. In this case, the effective Hamiltonians used in previous works, which describe the coupling between two Majorana end states only, do not apply. Thus, it is important to understand the robustness of the fractional Josephson effect when the Majorana end states are coupled to other fermions.

It has been recognized that the presence of other fermion states may not affect the function of the Majorana q-bit, as long as they are localized in the vicinity of the Majorana fermions. This is because as long as the temperature is much lower than the bulk band gap $\Delta_0$, thermal excitations of extended states are negligible and the total fermion parity, defined as the even or odd occupation of the fermions in the vicinity of the Majorana fermion, is conserved [\onlinecite{Akhmerov}]. However, in the context of the fractional Josephson effect, the following questions are raised. First, the fractional Josephson effect relies on the robustness of the avoided crossing of energy levels formed by the Majorana fermions as a function of $\phi$, the phase difference between the superconductors forming the junction. Is this avoided crossing robust in the presence of strong coupling to many fermion states? Second, what is the temperature dependence of the Josephson current? Third, while thermal excitation of the fermionic states preserves fermion parity, how does the difference of the fractional Josephson current which is proposed to measure the fermion parity depend on temperature? These are the questions we answer in this paper. 

\begin{figure}
\includegraphics[width=3.4in]{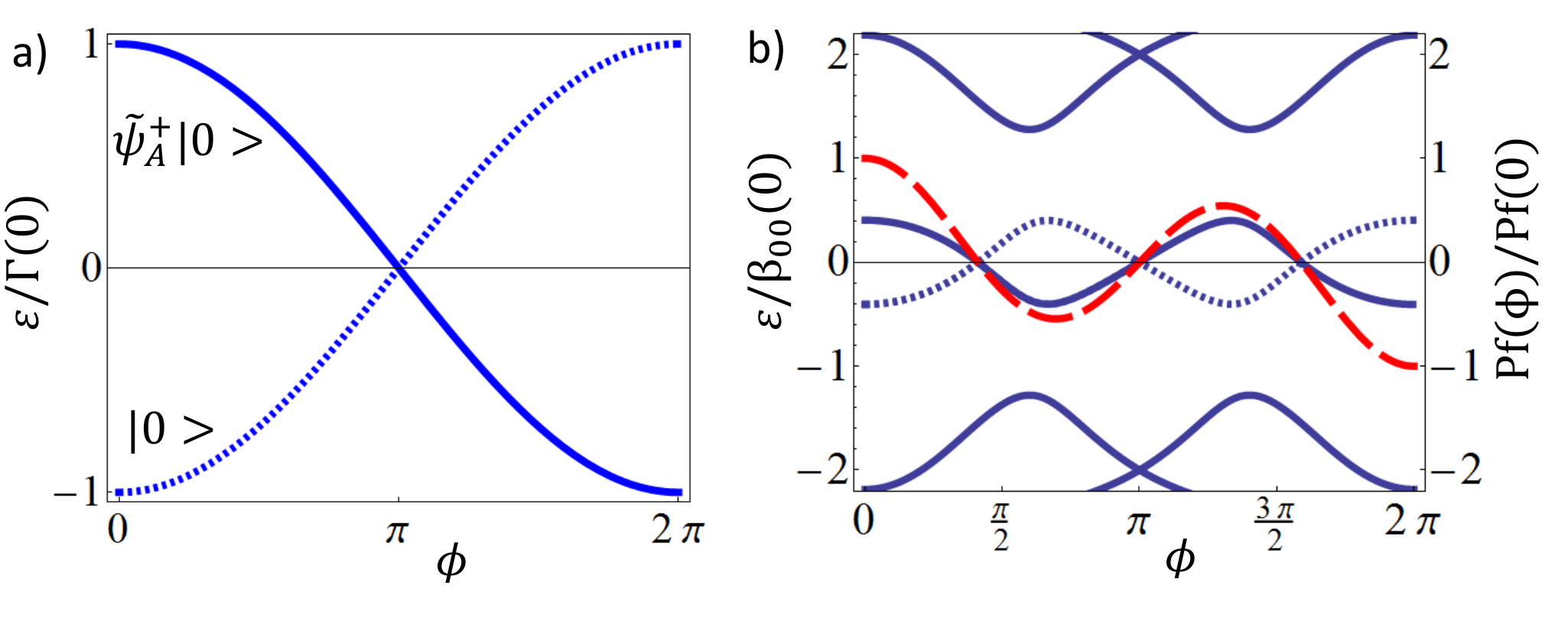}
\caption{\label{Fig2} a) The energy spectrum of the Andreev bound state when the state is occupied (solid line) and empty (dotted line). The crossing is protected by fermion parity conservation. b) The energy spectrum of $h_{M}$ (solid curves) and $\text{Pf}[h_{M}](\phi)/\text{Pf}[h_{M}](0)$ (dashed curve) as a function of $\phi$ are plotted. The presence of other fermions induces two more zero energy crossings. In this plot, every sign change of $\text{Pf}[h_M]$ is accompanied by an energy level crossing at zero energy. For the plot, $\gamma_{L/R,n}=2n$, $\beta_{n,n'}(\phi) \approx \cos(\phi/2)$ and $N=8$ are chosen in $H_{eff}$.}
\end{figure}

\emph{Review of the Single-channel Case}--- It was first pointed out by Kitaev [\onlinecite{Kitaev00}] that a single channel quantum wire with a proximity induced p-wave pairing gap can have Majorana end states localized at the ends of the quantum wire. When two superconducting wires with Majorana end states are joined together to form a Josephson junction, the two Majorana end states  $\eta_{L}$ and $\eta_R$ at the junction couple to each other. The effective Hamiltonian of the junction can be written as:
\begin{equation}
H_{1Deff}= 2i\Gamma\eta_{L} \eta_{R}=2\Gamma(\tilde{\psi}_{A}^{\dagger} \tilde{\psi}_{A}-\frac{1}{2}), \label{eff1}
\end{equation}
where $\Gamma$ is the coupling strength and $\tilde{\psi}_{A}=\eta_{L}+i\eta_{R}$ is the Andreev bound state formed by the two Majorana fermions. Suppose a gauge is chosen such that the superconducting phase  of the wire on the right is fixed, as the superconducting phase of the left quantum wire is increased by $2\pi$, all the fermions on the left of the junction acquire a minus sign such that $\eta_L \rightarrow -\eta_L$ and $\eta_R \rightarrow \eta_R$. Equivalently, the phase dependent can be absorbed into $\Gamma$. Thus, $\Gamma$ satisfies the condition:
\begin{equation}
\Gamma(\phi+2\pi)=-\Gamma(\phi).
\end{equation}
This condition shows that $\Gamma(\phi)$ is $4\pi$ periodic and crosses zero for certain $\phi=\phi_0$. For models in Refs.[\onlinecite{KSY,FKPRB09,SLTD}], $\Gamma(\phi)=\Gamma_0 \cos(\phi/2)$, where $\Gamma_0$ is a positive constant. The energy spectrum of the junction with $\Gamma(\phi)=\Gamma_0 \cos(\phi/2)$ is shown in Fig.\ref{Fig2}a.

Moreover, Eq.\ref{eff1} shows that $\Gamma(\phi)$ and $-\Gamma(\phi)$ are the eigenvalues of $H_{1Deff}$ and $\tilde{\psi}_{A}^{\dagger}|0>$ and $|0>$ are the corresponding eigenvectors, where $\tilde{\psi}_{A}|0>=0$. The energy of the occupied states $\tilde{\psi}_{A}^{\dagger}|0>$ and the empty state are shown by the solid and the dotted lines respectively in Fig.\ref{Fig2}a. If the fermion parity is conserved at the junction, this energy crossing is protected because $\tilde{\psi}_{A}^{\dagger}|0>$ and $|0>$ have different fermion parities and there is no transition from one state to the other when $\phi$ equals $2\pi$. Since the energy eigenvalues $\pm \Gamma(\phi)$ are $4\pi$ periodic and there are no transitions among the states with different fermion parity, the Josephson current, which is given by $I_{\pm}= \pm (2e/\hbar) \frac{d}{d\phi}\Gamma(\phi)$ is also $4\pi$ periodic. The observation of this $4\pi$ periodic Josephson current will be  strong evidence of the existence of Majorana fermions, and the sign of the current reveals the fermion parity of the junction.

\emph{Multi-channel Case}---In more realistic experimental situations, multiple transverse sub-bands of the quantum wires are occupied. It was pointed out in Ref.\onlinecite{PL} that Majorana end states exist in quasi-one dimensional wires when an odd number of bands are occupied. These Majorana end states are accompanied by fermionic end states with nonzero energy and the energy separation of the fermionic states is $\Delta_0/\sqrt{M}$ where $\Delta_0$ is the bulk superconducting gap and $M$ is the number of occupied transverse sub-bands. When two quasi-one dimensional wires are brought together to form a Josephson junction, the Majorana end states are coupled to each other and to the nonzero energy fermionic states as well. In this case, it is not obvious that the fractional Josephson effect is still robust. In this section, we show that the coupling of Majorana fermions with other fermionic end states may induce an additional even number of zero energy crossings, i.e. the total number of zero energy crossings is odd and the fractional Josephson effect survives, independent of the strengths of the couplings.

The effective Hamiltonian in the multi-band quantum wire case can be written as:
\begin{equation}
\begin{array}{l}
H_{eff}(\phi)= 2 i \beta_{00}(\phi) \Psi_{L,n=0} \Psi_{R,n=0} + \\ \sum^{N}_{n =1} [2\gamma_{Ln} (\Psi_{L,n}^{\dagger} \Psi_{L,n}-\frac{1}{2})+2\gamma_{Rn} (\Psi_{R,n}^{\dagger} \Psi_{R,n}-\frac{1}{2})] \\ +\sum'_{n,n'} \beta_{n,n'}(\phi) \Psi_{L,n}^{\dagger} \Psi_{R,n'}+h.c..
\end{array} \label{eff2}
\end{equation}
Here, $\Psi_{L/R,n=0}=\Psi_{L/R,n=0}^{\dagger}$ are Majorana end states from the left and right of the Josephson junction respectively. $\Psi_{L/R,n}$ with $n \neq 0$ are fermionic states with positive energy $\gamma_{L/R,n}$. Here, $N$ is the number of fermion end states which are within the bulk gap in each quantum wire, $\beta_{n,n'}(\phi)$ are the coupling strengths, $n$ and $n'$ in the second sum run from $0$ to $N$. The prime in the second sum indicates that the  $\beta_{00}(\phi)$ term is taken out from the sum and shown explicitly. As mentioned above, a phase change of the fermions on the left by $2\pi$ transforms $\Psi_{L}$ to $-\Psi_{L}$. This transformation is equivalent to transforming $\beta_{n,n'}$ to $-\beta_{n,n'}$. Thus, $\beta_{n,n'}$ satisfies the condition:
\begin{equation}
\begin{array}{cc}
\beta_{n,n'}(\phi+2\pi)=-\beta_{n,n'}(\phi).
\end{array}\label{coefficients1}
\end{equation}

Since the fermionic modes are coupled to the Majorana modes, it is convenient to use a basis which consists of Majorana modes only. Therefore, we write $\Psi_{L/R0}=\eta_{L/R0}$ and $\Psi_{L/R,n}= \eta_{L/R,n}+i\eta'_{L/R,n}$, $\Psi_{L/R,n}^{+}= \eta_{L/R,n}-i\eta'_{L/R,n}$ for the $n \neq 0$ modes, where $\eta$ and $\eta'$ are Majorana fermions. In this Majorana fermion basis, the Hamiltonian can be written as:
\begin{equation}
H_{M}= 
(\tilde{\eta_{L}}, \tilde{\eta_{R}}) h_{M} \left(
\begin{array}{c}
\tilde{\eta}_{L}^{T} \\ \tilde{\eta}_{R}^{T}
\end{array} \right)=
(\tilde{\eta_{L}}, \tilde{\eta_{R}}) \left(
\begin{array}{cc}
A & B \\ -B & C
\end{array} \right)
\left(
\begin{array}{c}
\tilde{\eta}_{L}^{T} \\ \tilde{\eta}_{R}^{T}
\end{array} \right) \label{Hm}
\end{equation},
where $\tilde{\eta_{a}} = (\eta_{a,0}, \eta_{a,1}, \eta'_{a,1} ...\eta_{a,N},\eta'_{a,N})$ with $a=L$ or $R$. $A$ and $C$ are $(2N+1) \times (2N+1)$ anti-symmetric Hermitian matrices, which describes the coupling between Majorana fermions on the same side of the junction. $B$ is a $(2N+1) \times (2N+1)$ matrix with arbitrary pure imaginary entries, which describe the coupling between Majorana fermions from opposite sides of the junction. Thus, $h_{M}$ is an $(4N+2) \times (4N+2)$ anti-symmetric Hermitian matrix.

According to Eq.\ref{coefficients1}, the sub-matrices satisfy the conditions:
\begin{equation}
\begin{array}{cc}
A(\phi+2\pi)=A(\phi), &  C(\phi+2\pi)=C(\phi) \\
\text{and} & B(\phi+2\pi)=-B(\phi).
\end{array}  \label{submatric}
\end{equation}
Therefore, changing the phase difference by $2 \pi$ is equivalent to applying a unitary transformation on $h_{M}$:
\begin{equation}
h_{M}(\phi+2\pi)=W^T h_{M}(\phi) W,
\end{equation}
where $W=\sigma_{z} \otimes I$ and $I$ is a $2N+1$ dimensional identity matrix. Using the fact that $\text{Pf}[W^T h_{M}(\phi) W]= \text{Pf}[h_M]\text{Det}[W]$ and $\text{Det}[W]=-1$, we have:
\begin{equation}
\text{Pf}[h_{M}(\phi+2 \pi)]= -\text{Pf}[h_M(\phi)], \label{Pf}
\end{equation}
where $\text{Pf}$ denotes the Pfaffian of an antisymmetric matrix. Since $\text{Pf}[h_{M}](\phi)$ is a continuous function of $\phi$ and $\text{Pf}[h_{M}](\phi)$ changes sign when the phase is increased by $2\pi$, $\text{Pf}[h_{M}](\phi)$ must cross zero an odd number of times when $\phi$ is advanced by $2 \pi$. From the relation $\text{Det}[h_{M}]=\text{Pf}[h_{M}]^2$, vanishing $\text{Pf}[h_{M}](\phi)$ indicates the appearance of zero energy modes.

Moreover, the eigenvalues of $h_{M}$ come in pairs of the form $\{ \epsilon_{i}(\phi), -\epsilon_{i}(\phi) \}$ because of particle hole symmetry, where $i$ runs from $1$ to $2N+1$ and the convention that $\epsilon_{i}>0$ at $\phi=0$ is chosen. As a result:
\begin{equation}
-i \text{Pf}[h_{M}](\phi)= \prod_{i}^{2N+1} \epsilon_{i}(\phi).   \label{sign}
\end{equation}

Suppose $i \text{Pf}[h_{M}](\phi)$ changes sign at $\phi = \phi_{0}$, Eq.\ref{sign} implies that an odd number of $\epsilon_{i}$ changes signs at $\phi_0$. Since $\text{Pf}[h_{M}](\phi)$ crosses zero an odd number of times and each crossing is accompanied by an odd number of energy level crossings at zero energy, thus, the total number of energy level crossings at zero energy must be odd. 

One example which illustrates the relation between energy level crossings and the sign of the Pfaffian is shown in Fig.\ref{Fig2}b. It is clear from Fig.\ref{Fig2}b that $\text{Pf}[h_{M}](2\pi)=-\text{Pf}[h_{M}](0)$ and every sign change of $\text{Pf}[h_M](\phi)$ is accompanied by a level crossing at zero energy. With an odd number of zero energy crossings, a phase change of $2\pi$ in a system described by the Hamiltonian $H_{M}$ does not change back to its original state. Instead, a phase change of $4\pi$ is needed. A physical consequence of this $4 \pi$ periodicity is the fractional Josephson effect. Therefore, the fractional Josephson effect is robust, even when the Majorana fermions are coupled to other fermions strongly. This is the first main result of this paper.

\emph{Finite Temperature Current}--- In order to demonstrate the non-Abelian properties of Majorana end states, the Majorana end states may undergo braiding operations using T-junctions suggested in Ref.\onlinecite{AOROF}. These braiding operations may result in changes in the fermion parity of the Josephson junction. In this section, we show that the Josephson current can be used to detect this change in fermion parity at the junction even at finite temperatures.

When there is only one single Andreev bound state at the junction and when the fermion parity is conserved, the Josephson current is 
\begin{equation}
I_{\pm}= \pm \frac{2e}{\hbar} \frac{d}{d\phi}\epsilon_{0}(\phi), \label{zeroTcurrent}
\end{equation}
where $\pm \epsilon_{0}(\phi)$ are the energy levels of the Andreev bound state when the Andreev bound state is occupied or empty respectively. In other words, the sign of the Josephson current gives the fermion parity of the junction. The Josephson current is temperature independent as long as thermal excitations to the extended fermionic states above the gap are negligible, so that fermion parity is conserved.

In the multi-channel case, if $\Delta_0 \gg k_BT \gtrsim \Delta_{0}/\sqrt{M}$, fermion parity is still conserved but the $n=0$ state can now exchange its occupation with the $n \neq 0$ fermion states. As a result, the Josephson current is temperature dependent. In order to calculate the Josephson current, we use the relation [\onlinecite{Beenakker}]:
\begin{equation}
I = \frac{2e}{\hbar}\frac{\partial}{\partial \phi} F, \label{current}
\end{equation}
where $F$ is the free energy of the Josephson junction. Suppose there are $N=\sqrt{M}$ fermionic Andreev bound states with energy $\epsilon_{n}>\epsilon_{0}$, where $\epsilon_0$ is the lowest positive energy state at $\phi=0$, the free energy can be written as:
\begin{equation}
F_{e/o} = -\frac{1}{\beta} \ln [\prod_{i>0}^{N}e^{\epsilon_i\beta}(e^{\epsilon_{0}\beta} P_{e/o}+e^{-\epsilon_{0}\beta} P_{o/e})].
\end{equation}
Here, fermion parity conservation at the Josephson junction is imposed such that the free energy with even and odd parity are denoted by $F_{e}$ and $F_{o}$ respectively. $P_{e}$ and $P_{o}$ denote the thermal weights of having an even or odd number of $\epsilon_{n\neq 0}$ states excited, which can be written as [ \onlinecite{remark1}]:
\begin{equation}
P_{e/o}= \sum_{k=even/odd}\frac{1}{k!}\frac{d}{dx^{k}}\prod_{i=1}^{N}(x+e^{-2\epsilon_{i}\beta})|_{x=0}. \label{Prob}
\end{equation}

The finite temperature current can be obtained from Eq.{\ref{current}} with two contributions to the fractional Josephson current. The Josephson current originated from the $\epsilon_0$ state:
\begin{equation}
I_{e/o}(\phi)= \frac{-2e}{\hbar} \big\{ \frac{e^{\epsilon_{0}\beta}P_{e/o}-e^{-\epsilon_{0}\beta}P_{o/e}}{e^{\epsilon_{0}\beta}P_{e/o}+e^{-\epsilon_{0}\beta}P_{o/e}}\big\} \frac{d}{d\phi}\epsilon_{0}(\phi) , \label{JC}
\end{equation}
and an extra contribution originated from the $\epsilon_{n \neq 0}$ states:
\begin{equation}
I'_{e/o}(\phi)= \frac{-2e}{\beta \hbar}\big\{ \frac{e^{\epsilon_{0}\beta} \frac{\partial}{\partial \phi} P_{e/o}-e^{-\epsilon_{0}\beta}\frac{\partial}{\partial \phi} P_{o/e}}{e^{\epsilon_{0}\beta}P_{e/o}+e^{-\epsilon_{0}\beta}P_{o/e}}\big\}.
\end{equation}

To extract the fractional Josephson current from a background $2\pi$ periodic Josephson current, we define the quantity:
\begin{equation}
\begin{array}{l}
\tilde{I}_{e/o}(\phi)=\frac{1}{2}[I_{Te/o}(\phi)-I_{Te/o}(\phi+2\pi)] \\ = \frac{1}{2}[I_{e/o}(\phi)-I_{e/o}(\phi+2\pi)]+\frac{1}{2}[I'_{e/o}(\phi)-I'_{e/o}(\phi+2\pi)]
\end{array}
\end{equation}
where $I_{Te}$ and $I_{To}$ denote the total Josephson current when the fermion parity of the junction is even and odd respectively. There are two contributions to $\tilde{I}_{e/o}(\phi)$. The contributions from the $\epsilon_0$ state and the $\epsilon_{n \neq 0}$ states are given by $I_{Fe/o}=\frac{1}{2}[I_{e/o}(\phi)-I_{e/o}(\phi+2\pi)]$ and $I'_{Fe/o}=\frac{1}{2}[I'_{e/o}(\phi)-I'_{e/o}(\phi+2\pi)]$, respectively. A nonzero value of $\tilde{I}(\phi)$ signals the presence of fractional Josephson effect. Moreover, using the fact that $\epsilon_{0}(\phi)=-\epsilon_{0}(\phi+2\pi)$ and $\epsilon_{n\neq0}(\phi)=\epsilon_{n \neq 0}(\phi+2\pi)$, we have:
\begin{equation}
\tilde{I}_{e}=-\tilde{I}_{o}.
\end{equation}
This is the second main result of our paper, which shows that the fractional Josephson current continues to depend on the fermion parity of the junction when the Majorana fermions are coupled to other fermionic modes and at finite temperatures.

\begin{figure}
\includegraphics[width=3.2in]{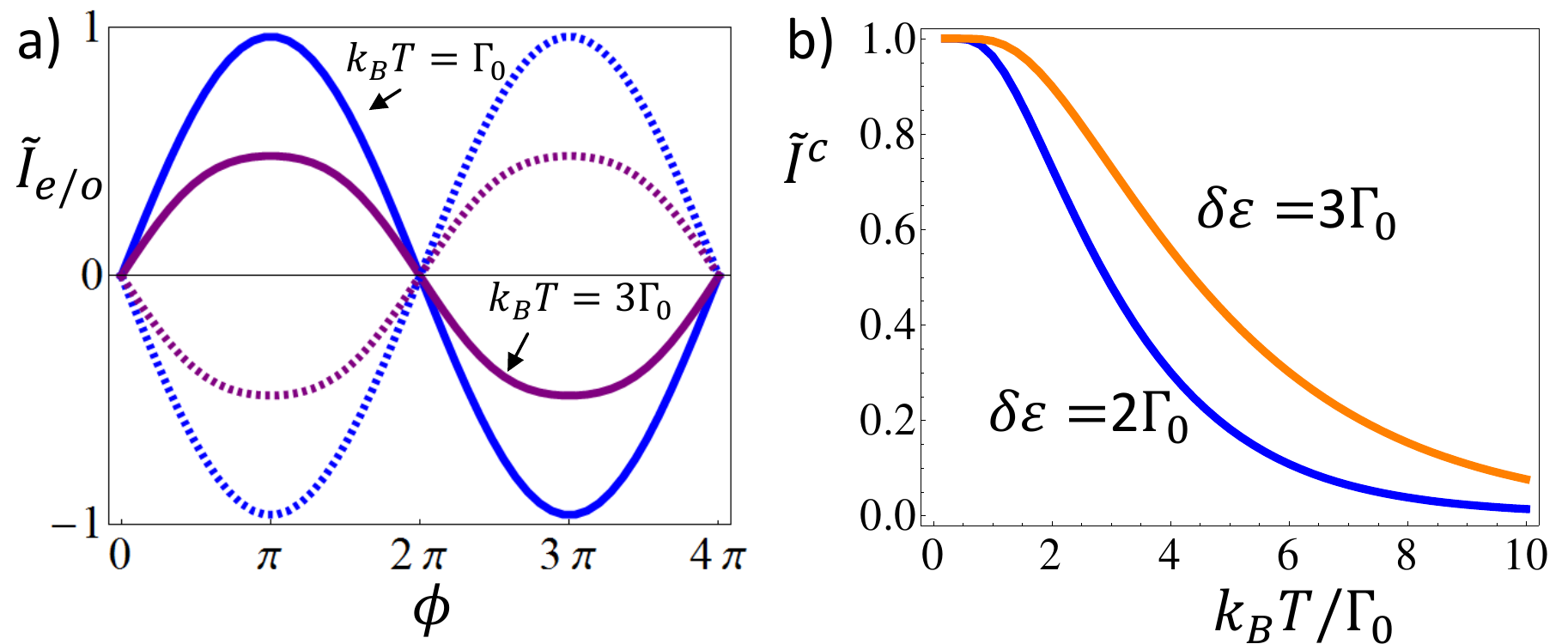}
\caption{\label{Fig3} a) $\tilde{I}_{e}$ and  $\tilde{I}_{o}$ verse $\phi$ at $\delta \epsilon = 2\Gamma_0$ and at temperatures $k_B T = 1 \Gamma_0$ and $k_B T = 3 \Gamma_0$.  Solid and dotted curves denote $\tilde{I}_{e}$ and $\tilde{I}_{o}$ respectively. $\tilde{I}_{e/o}$ are in units of $\frac{2e}{\hbar}\Gamma_0$. b) The maximal amplitude of the fractional Josephson current $\tilde{I}^c$ at $\delta \epsilon = 2\Gamma_0$ and $\delta \epsilon = 3\Gamma_0$ respectively.  }
\end{figure}

In the multi-band superconducting wire case, it is reasonable to consider the situation in which the two Majorana fermions $\eta_{L0}$ and $\eta_{R0}$ are weakly coupled to each other such that $\epsilon_0 \ll \delta \epsilon$, where $\delta \epsilon= \epsilon_{n+1}-\epsilon_{n}$ is the energy level spacing of the fermionic states. $\delta \epsilon$ is in the order of $\Delta_0/\sqrt{M}$ when the coupling between the fermionic end states are weak [\onlinecite{PL}]. In the regime where $\delta \epsilon \gg k_{B}T \approx \epsilon_{0}$, we have $P_e \approx 1 $ and $P_o \approx e^{-2\epsilon_{1}\beta} \ll 1$. In this case, $I_{Fe/o} \approx \mp \frac{2e}{\hbar}\frac{\partial}{\partial \phi}\epsilon_0$. On the other hand, $I'_{Fe/o} \approx \pm \frac{2e}{\hbar}(e^{-2\epsilon_0 \beta}-e^{2\epsilon_0 \beta})e^{-2\epsilon_{1} \beta} \frac{\partial}{\partial \phi}\epsilon_0$. Therefore, the $I'_{F}$ term is exponentially suppressed. The fractional Josephson current in this regime is given by the zero temperature result of Eq.\ref{zeroTcurrent}  multiplied by a temperature dependent prefactor which is exponentially small when $k_B T >  \epsilon_1$. The fractional Josephson current and the maximal amplitude of the current is plotted in Fig.\ref{Fig3}. A simple model with $\epsilon_{0}(\phi) = \Gamma_{0} \cos(\phi)$ and $\epsilon_{n}= \sqrt{(n\Delta_0/\sqrt{M})^2+\epsilon_0^2}$ is assumed. Fig.\ref{Fig3} shows that the maximal amplitude of the fractional current decreases exponentially as temperature increases. However, at temperature $k_B T \approx \delta \epsilon$, the fractional Josephson current is of order $ \frac{2e}{\hbar}\Gamma_0$. Assuming $\delta \epsilon = 2\Gamma_0 = 0.02meV$, we have $\tilde{I} \sim 1nA$ at $ T \approx 0.2K $.

Interestingly, in the regime where $ \epsilon_{0} \gg k_BT \gg \delta \epsilon$, the $I_{Fe/o}$ terms are suppressed and the $I'_{Fe/o}$ terms dominate. As a result, the $\phi$-dependent fractional Josephson current is very different from the ones in Fig.\ref{Fig3}a. This regime can be reached when the quasi-one dimensional quantum wires become very wide. The details of this work will be reported elsewhere.

In conclusion, we show that the absence of the avoided energy crossing, which gives rise to fractional Josephson effect, is robust even when Majorana fermions at the Josephson junction are coupled with arbitrary strengths to other fermion states. The fractional Josephson current can be used to measure the fermion parity of the junction at finite temperatures, as long as the temperature does not greatly exceed the excitation energy to the other fermions.

\emph{Acknowledgments}--- We thank K.T. Chen, L. Fu, T.K. Ng and A.C. Potter for insightful discussions. KTL acknowledges the support of Croucher Foundation, PAL acknowledges the support of DOE grant number DE-F602-03ER46076 and the hospitality of IAS-HKUST.


\begin{thebibliography}{99}
\bibitem{remark1} $N$ even is assumed in the text. When $N$ is odd, the sum of $k$ in Eq.\ref{Prob} is changed to $\sum_{k=odd/even}$.
\end{thebibliography}
\end{document}